\documentclass{article}

\usepackage{PRIMEarxiv}

\usepackage[utf8]{inputenc} % allow utf-8 input
\usepackage[T1]{fontenc}    % use 8-bit T1 fonts
\usepackage{hyperref}       % hyperlinks
\usepackage{url}            % simple URL typesetting
\usepackage{booktabs}       % professional-quality tables
\usepackage{amsfonts}       % blackboard math symbols
\usepackage{nicefrac}       % compact symbols for 1/2, etc.
\usepackage{microtype}      % microtypography
\usepackage{lipsum}
\usepackage{fancyhdr}       % header
\usepackage{graphicx}       % graphics
\graphicspath{{media/}}     % organize your images and other figures under media/ folder

\usepackage{graphicx}% Include figure files
\usepackage{dcolumn}% Align table columns on decimal point
\usepackage{bm}% bold math
\usepackage{amsmath,amssymb}
\usepackage{xcolor}
\usepackage{longtable}
\usepackage[longtable]{multirow}

\newcommand{\me}{{\rm{e}}}
\newcommand{\mi}{{\rm{i}}}

\title{Monolithic atom interferometry
}

\author{
 Johannes Fiedler\\
Department of Physics and Technology\\ University of Bergen, \\ Bergen, Norway\\
\texttt{johannes.fiedler@uib.no} \\
   \And
  Kim Lefmann\\
Niels Bohr Institute\\
University of Copenhagen\\  K{\o}benhavn, Denmark \\
\And
Wolf von Klitzing\\
Institute of Electronic Structure and Laser\\ Foundation for Research and Technology-Hellas\\
Heraklion, Greece\\
\And
 Bodil Holst\\
Department of Physics and Technology\\ University of Bergen, \\ Bergen, Norway\\
\texttt{bodil.holst@uib.no}
}

\begin{document}
\maketitle

\begin{abstract}
Atom and, more recently, molecule interferometers are used in fundamental research and industrial applications. Most atom interferometers rely on gratings made from laser beams, which can provide high precision but cannot reach very short wavelengths and require complex laser systems to function. Contrary to this, simple monolithic interferometers cut from single crystals offer (sub) nano-meter wavelengths with an extreme level of stability and robustness. Such devices have been conceived and demonstrated several decades ago for neutrons and electrons. Here, we propose a monolithic design for a thermal-beam molecule interferometer based on (quantum) reflection. We show, as an example, how a reflective, monolithic interferometer (Mach-Zehnder type) can be realised for a helium beam using Si(111)-H(1$\times$1) surfaces, which have previously been demonstrated to act as very robust and stable diffractive mirrors for neutral helium atoms.
\end{abstract}

% keywords can be removed

\section{\label{sec:level1}Introduction}
The field of atom interferometry has expanded enormously over the last few decades. Atom interferometers are used in various applications, from magnetic and gravity sensing~\cite{stray2022quantum, hardman2016simultaneous}, quantum metrology~\cite{riedel2010atom-chip-based} to atomic clocks~\cite{ludlow2015optical}. They may even be used as dark matter and gravitational wave detectors~\cite{canuel2020elgar} also in space~\cite{el-neaj2020aedge, Tino2013}. Compact, portable atom gravimeters for prospecting, oil survey and geophysical investigations have recently become commercially available~\cite{desruelle2018}. Atom interferometers will also be useful as accelerometers for sub-sea navigation in submarines and, more recently, underwater drones~\cite{app10041256,hardman2016simultaneous}. This, however, will require very compact solutions, which are not presently available.

Atom interferometers use either cold atoms (including Bose-Einstein Condensates)~\cite{hall1998measurements} or thermal atoms beams~\cite{Pritchard1991}, and more recently hot thermal  vapours~\cite{biedermann2017atom}. Most optical interferometers have, by now, been realised as atom interferometers, including Young's double slit, Mach-Zehnder, Talbot-Lau,  Ramsey-Bord\'e and Sagnac interferometers. 

Historically, Young's double slit makes the simplest atom interferometer. The beam is split into two paths by passing through a double slit, and the interference pattern is observed on a screen further down the beam path. It was realised for atoms for the first time in 1991 using metastable helium atoms passing through a thin gold foil~\cite{Carnal1991}.

The simplest split-path atom interferometer is arguably the Mach-Zehnder interferometer. It exploits  the de Broglie wavelength of the atoms in a diffraction grating configuration with split beam paths. The first Mach-Zehnder atom interferometer was realised in 1991~\cite{Pritchard1991} using a sodium beam and solid transmission diffraction gratings. Later in 1995, it was developed further by using metastable neon and argon and transmission diffraction gratings made of standing light waves~\cite{Lee1995,Zeilinger1995}, in 2002 using ground-state lithium also with light-wave gratings~\cite{Vigue2002} and later again using neutral helium with solid gratings. Results from the last mentioned instrument were never published,  but it is mentioned in a review paper from 2009~\cite{Pritchard2009}.

In the Talbot-Lau interferometer, the self-imaging property of a grating is exploited in near-field diffraction. The atom paths are not truly separated; therefore, this type of interferometer has been used extensively for experiments with heavy molecules where the de-Broglie wavelength is very small. The first Talbot-Lau atom interferometer was realised in 1994~\cite{Li94}. 

Where the Mach-Zehnder interferometer and the Talbot-Lau interferometers are adapted from light optics, the Ramsey-Bord\'e interferometer, first realised in 1949 by Norman Ramsey~\cite{PhysRev.78.695}, can only be used for atoms: the principle is diffraction by absorption of a single photon on a weakly allowed transition to split the wave package. In the 1980th, this interferometer has been further developed by Christian Bord\'e by using atomic recoil to create a beam splitter~\cite{PhysRevA.30.1836}. This interferometer type is currently the standard for high-precision measurements, such as atomic clocks.

In light optics, the Sagnac interferometer, also called ring interferometer, relies on a beamsplitter mirror to create two beams that travel equidistant paths in opposite directions through a ring structure guided by reflective mirrors. The two beams meet at the starting point, where they interfere and are made to exit the ring. The first atom interferometer using the Sagnac effect was realised in 1991 using a Ramsey–Bordé configuration of a state-labelled atom interferometer based on single-photon transitions, with a beam of atoms traversing two pairs of travelling wave fields. The laser fields within each pair are separated by a distance $D$, while the two pairs are separated by $d$ and are counter-propagating with respect to each other~\cite{Riehle91}. By rotating the interferometer, the counter-propagating beams collect different phases along their optical paths leading to an interference pattern on the screen. Such a configuration provides an absolute measurement of the rotational speed.

The atomic structure of a single crystal offers a simple periodic diffractive grating. Thus, it could produce many different types of interferometers, where the monolithic construction guarantees extreme stability. Interferometers based on transmission through solid slabs of material have been demonstrated, e.g.  X-rays~\cite{doi:10.1063/1.1754212,Chetwynd_1991}, neutrons~\cite{Rauch2015} and electrons~\cite{berggren2017}. Unfortunately, these techniques are inapplicable to atoms, which interact too strongly with any solid material they travel through. Monolithic atom interferometers have been used widely in neutron scattering experiments observing gravitationally induced interference (in transmission)~\cite{PhysRevLett.34.1472} and the quantised states of neutrons in the presence of gravitational fields with perfectly reflecting mirrors~\cite{Nesvizhevsky2002}. Neutrons are sensitive to external forces and, thus, suitable candidates for quantum sensing. However, such experiments require an extensive, costly infrastructure to create, control and detect the neutron beam. This also applies to cold atom interferometers. Thermal atom beams are easier to create and couple more robust to external fields due to the higher mass of the atoms. A further advantage of thermal atom interferometers is that they can operate continuously, dramatically decreasing the temporal resolutions.

\begin{figure}[t]
    \centering
    \includegraphics[width=0.4\columnwidth]{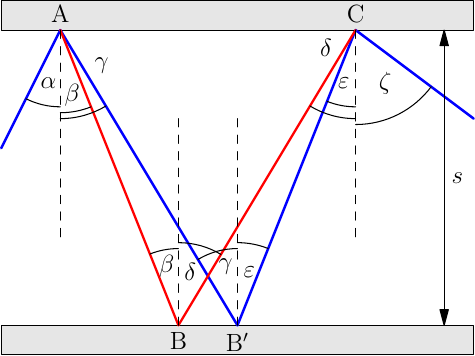}
    \caption{Sketch of the optical paths within a monolithic, reflection interferometer: the beam is reflected three times between the surfaces of two parallel slabs (grey area) separated by the distance $s$. The incoming beam is reflected at point A with an incidence angle $\alpha$. Two different diffraction orders are selected: Reflection towards point B with diffraction angle $\beta$  and reflection towards B${}'$ with diffraction angle $\gamma$. At point B, the incidence angle is the same as the outgoing angle in point A: $\beta$. Part of this beam is  reflected towards point C with diffraction angle $\delta$. At point B${}'$, the incidence angle is given by $\gamma$ due to the reflection at point A and reflected at the diffraction angle $\varepsilon$. In point C, the incoming waves with incidence angles $\delta$ and $\varepsilon$ are recombined, leaving the slab with a reflection angle $\zeta$.}
    \label{fig:schematic}
\end{figure}

Here, we propose a novel interferometer based on the reflection of atoms on monolithic single-crystal structures.  The basic operation principle is depicted in Fig.~\ref{fig:schematic}: an incident beam of atoms is reflected by the crystal lattice (A) into two components, which impinge onto a second mirror and recombine on the third reflection.

In the past, atoms had been neglected, largely because the atoms most commonly used in interferometry  (Rubidium, Rb~\cite{PhysRevLett.115.013004}; Caesium, Cs~\cite{PhysRevLett.78.2046}; Argon, Ar~\cite{doi:10.1080/09500349708231906}; Sodium, Na~\cite{PhysRevLett.66.2693}; Potassium, K~\cite{PhysRevA.49.R2213}) will stick to surfaces under most conditions. Similarly, metastable atoms, which have also been used for interferometry (Argon, Ar~\cite{PhysRevLett.127.170402}; Helium, He~\cite{Carnal1991}), will decay upon impingement. A further practical challenge for a reflection-based interferometer is the contamination of the reflecting surface, which distorts the diffraction. For example, all metal surfaces will be covered in physisorbed molecules within hours, even in ultra-high vacuum~\cite{doi:10.1080/01442350500037521,doi:10.1146/annurev-physchem-040214-121958,saxena1989thermal}.

Noble gasses, including ground-state helium, H$_2$, HCl and  other molecules,  are known to scatter from various surfaces over a broad temperature range without sticking to them~\cite{D2CP03369K,doi:10.1063/5.0058789,doi:10.1063/5.0026228}. Over the last years,  focusing mirrors for neutral, ground-state helium have been developed for neutral helium microscopes~\cite{palau2021}. An important requirement for these mirrors is that they must remain stable in a vacuum for months. One of the solutions implemented was Si(111)-H$(1 \times 1)$~\cite{holst1997}. Detailed experiments on He and H$_2$ scattering were performed~\cite{BARREDO200724,doi:10.1063/1.480723} and the interaction potential between Helium and Si(111)-H$(1 \times 1)$ calculated~\cite{doi:10.1063/1.480723}. This interaction potential was then used to obtain the intensity of the different diffraction peaks for a range of conditions~\cite{Buckland1999}.

The advantage of the Si(111)-H$(1 \times 1)$ surface from an experimental point of view is that it can be prepared chemically by dipping the Si(111) crystal in an HF solution~\cite{MACLAREN2001285}. This means a monolithic configuration with two reflecting surfaces facing each other can be fabricated at any spacing. The additional advantage of using the Si(111)-H$(1 \times 1)$ surface is the small lattice constant of $a_{\rm S}=3.383$~\AA~\cite{ BARREDO200724} which, together with the wavelength of, as an example, helium atoms in a room temperature beam: $\lambda_{\rm dB} =0.55\,\rm{\AA}$, ensures a very big wave-package separation. Recent matter-wave interferometers typically split the wave package over a few milliradian~\cite{Brand2015,PhysRevLett.125.033604,PhysRevLett.127.170402,Bruhl_2002}. In contrast, using the room-temperature helium beam described above, the proposed new interferometer splits the matter wave over 0.5~radians.

The atom interferometer we introduce here uses reflective atom-surface diffraction as a beam splitter. Further reflections from a parallel surface yield the recombination of the wave and thus the interference; see Fig.\,\ref{fig:schematic}. We present a theoretical model determining the expected interference patterns and apply the model to the interference of helium atoms using Si(111)-H$(1 \times 1)$ surfaces, where we concentrate on describing the general principles by describing an ideal system with a perfectly coherent and monochromatic beam and an experimentally based model for the diffraction probabilities. We have chosen an experimentally realisable parameter set providing all possible superpositions occurring in such interferometer: single-path transmission, double-path superposition with vanishing phases and multipath interference. Finally, we discuss how a reflective interferometer based on quantum reflection can be addressed. The paper finishes with a conclusion and outlook on future work.

\section{The reflective interferometer}\label{sec:inter}

\subsection{Geometric arrangement}\label{sec:reflinter}

The general arrangement of a monolithic reflection interferometer is depicted in Fig.~\ref{fig:schematic}. A slab is cut into a U-shaped monolith to form two parallel planar surfaces with a distance $s$ being sufficiently large to achieve propagating waves inside the interferometer. A particle beam will be diffracted minimally three times at points A, B and C. The beam will be split at point A, and each part will be reflected at point B and recombined in point C, where they interfere.

In detail: a particle beam is sent via an incidence angle $\alpha$ towards one surface. It is reflectively split in point A into a range of diffraction orders determined by the incident beam angle $\alpha$,  the periodic surface structure described by the lattice spacing $a_{\rm S}$, and the beam wavelength $\lambda$ through the well known reciprocal lattice equation~\cite{Kittel2004}. We pick two orders, the first one with the reflection angle $\beta$
\begin{equation}
\sin\beta = \sin \alpha +\frac{n_1\lambda}{a_{\rm S}}\,, \label{eq:diff}
\end{equation}
with an integer $n_i\in\mathbb{Z}$ (numerating the diffraction order), and the second one with reflection angle $\gamma$
\begin{equation}
\sin\gamma = \sin \alpha +\frac{n_{1'}\lambda}{a_{\rm S}}\,. 
\end{equation}
At point A, the two selected diffraction orders propagate towards points B and B${}'$, where they are reflected towards point C and recombine. Point B denotes the reflection point one diffraction order from point A; thus, the corresponding incidence angle is $\beta$. To satisfy the recombination of the beam, the reflection angle $\delta$ has to be of a non-zeroth diffraction order expressed as
\begin{equation}
    \sin\delta= \sin \beta + \frac{n_2\lambda}{a_{\rm S}} = \sin\alpha +\frac{(n_1+n_2)\lambda}{a_{\rm S}}\,.
\end{equation}
Analogously, the reflection at point B${}'$ can be determined by
\begin{equation}
    \sin\varepsilon = \sin\gamma + \frac{n_{2'}\lambda}{a_{\rm S}}=\cos \alpha + \frac{\left(n_{1'}+n_{2'}\right)\lambda}{a_{\rm S}}\,.
\end{equation}
To satisfy the recombination of the beam at point C, the diffraction of the incoming beams need to occur under the same diffraction angle, which can be described mathematically by the relation
\begin{equation}
    \sin\zeta = \sin\delta + \frac{n_3\lambda}{a_{\rm S}} = \sin\alpha  + \frac{\left(n_1+n_2+n_3\right)\lambda}{a_{\rm S}}\,,
\end{equation}
and 
\begin{equation}
\sin\zeta = \sin\varepsilon  + \frac{n_{3'}\lambda}{a_{\rm S}}= \sin\alpha  + \frac{\left(n_{1'}+n_{2'}+n_{3'}\right)\lambda}{a_{\rm S}}\,.
\end{equation}
These equations yield a constrain for the diffraction orders
\begin{equation}
    n_{3'} = n_1+n_2+n_3-n_{1'}-n_{2'} \,.\label{eq:condition1}
\end{equation}
In addition to this angular dependence, the distance between points A and C needs to be the same for both paths to satisfy the recombination of the beams. Figure~\ref{fig:schematic} illustrates this condition: the blue and red beamlines need to recombine in the same point C. Otherwise, they would be reflected without any spatial overlap to interfere directly. If they are reflected into parallel beams from different spots, they will interfere in the far field with a phase shift proportional to the spatial difference between both points. To achieve interference also in the  optical near-field regime for the entire interferometer, the condition reads
\begin{equation}
    \tan\beta+\tan\delta = \tan\gamma+\tan\varepsilon \,.\label{eq:condition2}
\end{equation}
Finally, we sum up six parameters characterising a reflective atom interferometer which have to satisfy the conditions~(\ref{eq:condition1}) and (\ref{eq:condition2}). The latter can either be used for determining the incidence angle $\alpha$ or by rewriting the equation
\begin{eqnarray}
\tan\left(c+N_1\right) +\tan\left(c+N_1+N_2\right)- \tan\left(c+N_{1'}\right)-\tan\left(c+N_{1'}+N_{2'}\right) =0 \,,   
\end{eqnarray}
with $c=\cos\alpha$ and $N_i = n_i\lambda/a_{\rm S}$, one finds the following conditions leading to an $\alpha$-independent solution:
\begin{eqnarray}
n_1 = n_{1'} + n_{2'} \wedge n_{1'} = n_1 +n_2 \,.
\end{eqnarray}
The interference pattern is due to the phase shift along the different optical paths ABC and AB${}'$C. The path lengths can be determined via these angles for the path along point B
\begin{equation}
    b=s\left(\frac{1}{\cos\beta}+\frac{1}{\cos\delta}\right)\,, \label{eq:b}
\end{equation}
and along the point B${}'$
\begin{equation}
    b'=s\left(\frac{1}{\cos\gamma}+\frac{1}{\cos\varepsilon}\right)\,.\label{eq:bs}
\end{equation}
The interference occurs via the superposition of two waves with the same wave vector ${\bm{k}}$, but are phase shifted with respect to the respective path lengths, $b-b'$. Hence, the phase shifts between the different paths are given by
\begin{equation}
    \varphi = k(b-b') \,.\label{eq:phase}
\end{equation}
It can be observed in Eqs.~(\ref{eq:b}) and (\ref{eq:bs}) that the path lengths are proportional to the slab separation $s$ and, thus, $s$ should be tuned with respect to the wave vector to maximise the phase shift between both interfering beams.

\begin{figure}[t]
    \centering
    \includegraphics[width=0.6\columnwidth]{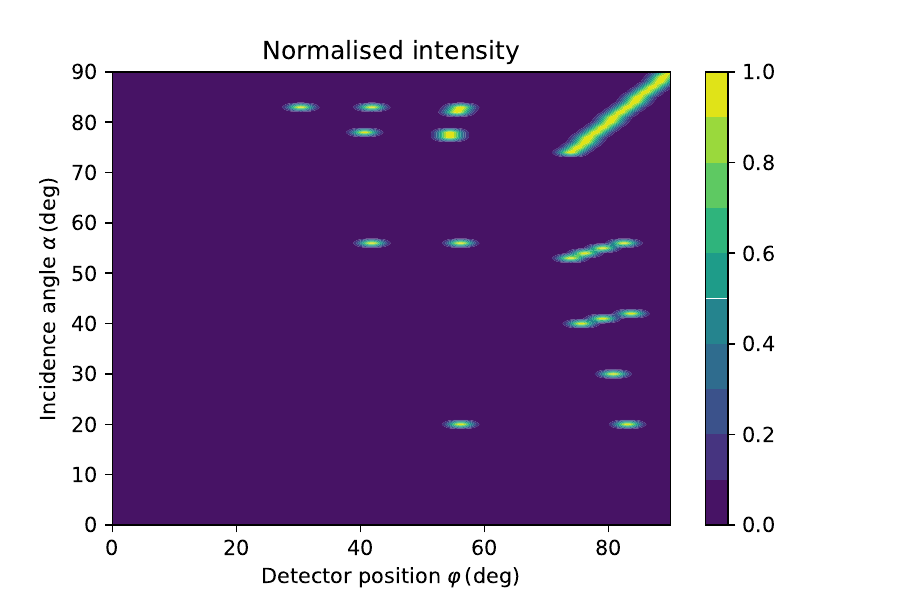}
    \caption{Distribution of the diffraction orders on the screen $\varphi$ depending on the incidence angle $\alpha$ for a monolithic atom interferometer built of silicon with hydrogen passivised surfaces, which are separated by 5~mm and have an extension of 50~mm. The considered wavelength was $\lambda = 0.55\,\text{\AA}$. The purple area describes the dark regions where no particle will appear. For each incidence angle $\alpha$, the maximum population of the diffraction order is marked in yellow. The remaining peak intensities are plotted relative to the maximum intensity according to the colour scale.}
    \label{fig:diffractionorders}
\end{figure}

Figure~\ref{fig:diffractionorders} illustrates the positions of the different diffraction for different incidence angles $\alpha$ for a particular interferometer configuration. It can be seen that the diffraction orders are strongly separated. All lines are discontinued due to the finite length of the interferometer, which leads to some beams escaping the interferometer. These particles will likely hit the surface and fall into the interferometer; thus, they will not affect the interference patterns.

\begin{figure*}[t]
    \centering
    \includegraphics[width=0.8\textwidth]{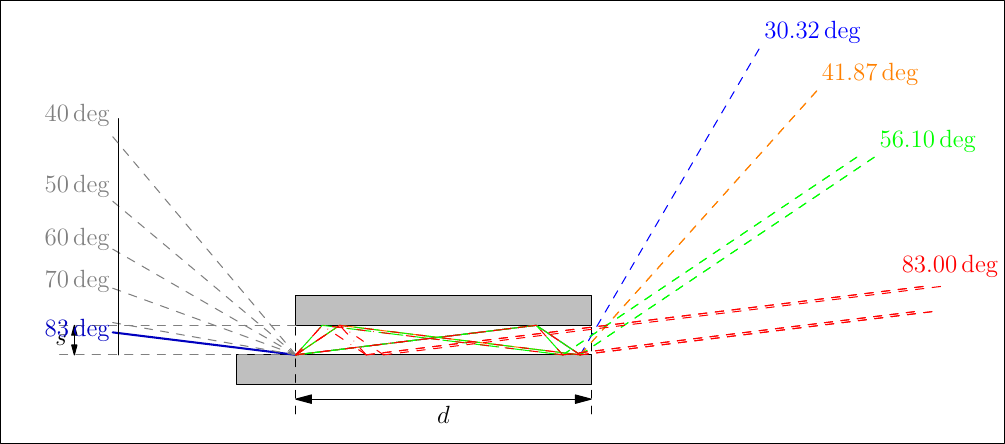}
    \caption{Optical paths in a monolithic reflective atom interferometer: a slab cut into a monolithic crystal of length $d$ (50~mm) and width $s$ (5~mm). A Helium beam with an incidence angle of $83$ deg (dark blue line) enters the interferometer. It is diffracted at the hydrogen passivised surfaces with a lattice constant $a_{\rm S} =3.383\,\rm{\AA}$. The diffracted orders are reflected two more times until they leave the interferometer. It can be seen that the third-order ($-3 =n_1+n_2+n_3$) diffraction beam will not show any interference  (blue dashed line at $30.32$ deg), see table~\ref{tab:diff}, the second-order beam at $41.87$ deg (orange lines) will not show any interference due to equal optical path lengths; the diffraction at $56.10$ deg (green lines), the zeroth order at $83.00$ deg (red lines) will be measured separately in the near-field regime, whereas they will interfere in the far-field leading to the interference patterns depicted in Fig.~\ref{fig:diff}.}
    \label{fig:Interferometer}
\end{figure*}
\subsection{Reflection coefficients for the different beam paths inside the interferometer}
In the last section, the conditions for interference were obtained. We now consider the intensity distribution in the interference signal, described via a reflection function.
This reflection function depends on the incidence and diffraction angle $\vartheta_1$ and $\vartheta_2$, respectively. We model the reflected beam via a Gaussian intensity distribution. Consequently, each diffraction order has a Gaussian profile which we normalise to the real-valued probability of each diffraction order $\rho_n$
\begin{equation}
    r(\vartheta_1,\vartheta_2) = \sum_n \rho_n\me^{-\frac{\left(\vartheta_2 - \theta_n\right)^2}{2\sigma_n^2}}\,,\label{eq:refl}
\end{equation}
with the width of the diffracted signal $\sigma_n$ and the position of the diffracted beam $\theta_n$ determined by Eq.~(\ref{eq:diff}). The widths depend on the incidence angle and wavelength, $\sigma_n=\sigma_n (\lambda,\vartheta_1)$. These impacts are negligible for surface diffraction, the manuscript's content, due to the overall weak reflection signal~\cite{doi:10.1063/1.480723}. The reflection coefficient~(\ref{eq:refl}) only includes the inelastic scattering, that the wavelength of the outgoing wave is the same as that of the incoming wave, $\lambda_{\rm inc} = \lambda_{\rm out}$. Thus, the total reflected signal is smaller than one, $\int\mathrm d \vartheta_2\, r(\vartheta_1,\vartheta_2) <1$.

In general, there are five lengths involved in such an interferometer: the wavelength ($\lambda_{\rm dB}$), the dimensions of the interferometer (length $d$ and slap separation $s$) and the free-space propagation lengths (source to interferometer $L_1$ and interferometer to detector $L_2$). Typically, these dimensions are on different length scales $\lambda_{\rm dB} \ll d,s < L_1, L_2$. This consideration allows for the separation of length scales. Consequently, each particle will only interfere with itself inside the same optical path in the interferometer. Thus, we can treat each path inside the interferometer separately, and the collected diffraction image will follow from the Gaussian beam envelope. The partial waves will experience a different phase shift due to the optical path~(\ref{eq:phase}). Thus, we can describe the reflection properties of the entire \textit{inter}ferometer with a single modified reflection coefficient
\begin{align}
    r_{\rm inter}(\vartheta_1,\vartheta_2) = \sum_{n_1n_2n_3}\rho_{n_1} \rho_{n_2} \rho_{n_3} f_{n_1n_2n_3} \me^{\mi k b_{n_1n_2}}\me^{-\frac{\left[\vartheta_2 - \theta_{n_1n_2n_3}(\vartheta_1)\right]^2}{2\sigma^2}}\,,
\end{align}
with the wave vector of the matter wave, $k = 2\pi/\lambda$, and the indicator function $f_{n_1n_2n_3}$ factoring in the interferometer's geometry (which determines whether the beam can pass through the interferometer or not). The beam spread of all diffraction orders will usually be the same for a monochromatic wave, $\sigma=\sigma_n$ for all diffraction orders $n$. Due to the tilted reflective surfaces with respect to the beam incidence, the detected spots will be slightly asymmetric, which we neglect for consideration in this manuscript. The position of the diffraction order is given by $\theta_{n_1n_2n_3}(\vartheta_1)$ which is the three-times composition of Eq.~(\ref{eq:diff}) simplifying to
\begin{equation}
    \theta_{n_1n_2n_3}(\vartheta_1) = \arcsin\left[\sin\vartheta_1 +\frac{\left(n_1+n_2+n_3\right)\lambda}{a_{\rm S}}\right] \,.
\end{equation}

\subsection{The monolithic interferometer for He and Si(111)-H$(1 \times 1)$}
Let us consider a helium beam with de-Broglie wavelength $\lambda_{\rm dB} = 0.55$\AA\, and a beam spread of 1~mrad at a distance of 1~m from the interferometer (propagation length $L_1=1\,\rm{m}$). This corresponds to a 1~mm beam waist ($w= 1\,\rm{mm}$) as the beam enters the interferometer. The lattice spacing of Si(111)-H(1$\times$1) is $a_{\rm S} = 3.383$~\AA~\cite{BARREDO200724}. We consider an incidence angle of 83 deg and the reflection coefficient~(\ref{eq:refl}) with the amplitudes $\rho_0=0.06$, $\rho_{\pm1} = 0.03$ and $\rho_{\pm2} =0.015$. These scattering values correspond to the experimentally obtained data for a beam with a wavelength of $0.6$ \AA\, and an incidence angle of $52$ deg reported in Ref.~\cite{BARREDO200724}. We changed the incidence angle because the result would be restricted to the zeroth order; see Fig.~\ref{fig:diffractionorders}. Table~\ref{tab:diff} shows the ratio of transmitted atoms into each  diffraction channel. The reflection coefficients influence only the amplitude of the interference patterns, not the position of the peaks. Thus, the impact of the correct scattering amplitudes is neglectable. We have chosen the parameters to demonstrate several effects: the single-beam transmission, the two-path superposition and the multi-path interference. Here we restrict our considerations to the zeroth, first and second diffraction orders. Furthermore, we consider the reflecting plates to be 50~mm long and 5~mm separated from each other. The optical paths for this scenario are depicted in Fig.~\ref{fig:Interferometer}. It can be seen that the third-order diffraction beam (at $30.32$ deg, blue line) consists of a single beam. It thus will not show any interference; the second-order beam (at $41.87$ deg, orange line) is the superposition of two paths, as described in Sec.~\ref{sec:reflinter}, but with  equal optical paths which again will not interfere; and the first and zeroth order will show two separate signals each, that will lead to interference in the far field.

\begin{table}[t]
    \centering
    \begin{tabular}{c|c|c|c|c|c|c}
    Angle $\varphi\,(\rm{deg})$ & $n_1$ & $\beta(n_1)$ & $n_2$ & $n_3$ & path $b \,(\rm{cm})$ & Ampl. $a$ (\%)\\\hline
30.32 & 0 &1.4486 & -1  & -2 & 5.00 & 0.0270\\
41.87 & -1 &0.9791 & 1 & -2 & 5.00 & 0.0135\\
41.87 & 0 &1.4486 & -1 & -1 & 5.00 & 0.0540\\
56.10 & -2 &0.7307 & 2 & -1 & 4.77 & 0.0068\\
56.10 & 0 &1.4486 & -2 & 1 & 4.77 & 0.0270\\
56.10 & -1 &0.9791 & 1 & -1 & 5.00 & 0.0270\\
56.10 & 0 &1.4486 & -1 & 0 & 5.00 & 0.1080\\
83.00 & -2 &0.7307 & 1 & 1 & 1.57 & 0.0135\\
83.00 & -2 &0.7307 & 2 & 0 & 4.77 & 0.0135\\
83.00 & 0 &1.4486 & -2 & 2 & 4.77 & 0.0135\\
83.00 & -1 &0.9791 & 0 & 1 & 1.79 & 0.0540\\
83.00 & -1 &0.9791 & 1 & 0 & 5.00 & 0.0540\\
83.00 & 0 &1.4486 & -1 & 1 & 5.00 & 0.0540\\
83.00 & -1 &0.9791 & -1 & 2 & 1.57 & 0.0135
\end{tabular}
    \caption{Overview of beams expected from an 83 deg incidence angle reflected into the diffraction angle $\varphi$ with the diffraction orders $n_1$, $n_2$ and $n_3$ with the first diffraction angle $\beta$ in radians. The last columns describe the optical path length $b$ and amplitudes $a$ being the ratio of transmitted atoms into each diffraction channel (\%).}
    \label{tab:diff}
\end{table}

\begin{figure}[t]
    \centering
    \includegraphics[width=0.6\columnwidth]{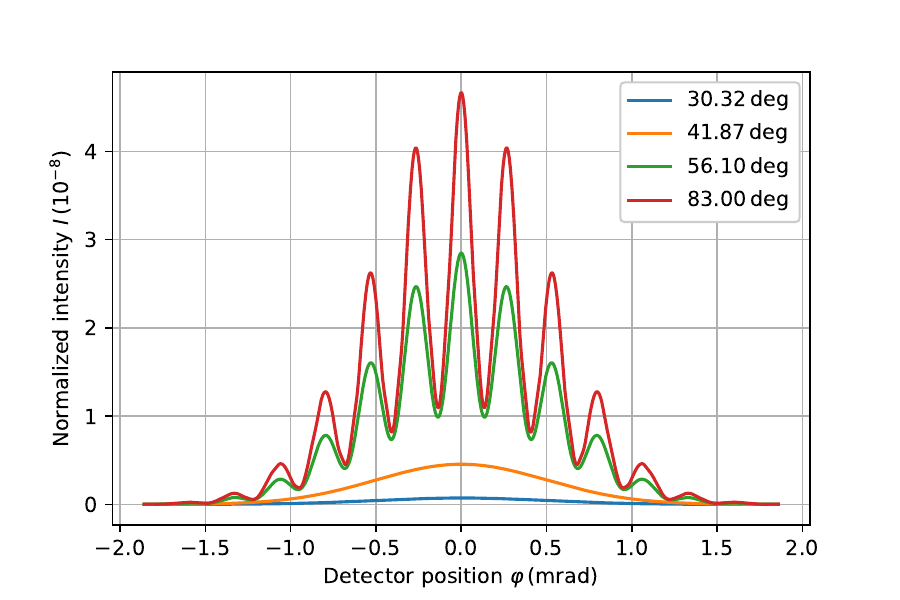}
    \caption{Far-field diffraction patterns of each spot at 30.32 deg (blue line), 41.87 deg (orange line), 56.10 deg (green line) and 83.00 deg (red line) for a Helium beam with wavelength $\lambda_{\rm dB} = 0.55\,\rm{\AA}$ with an incidence angle of 83 deg. The vertical black dashed line marks the regime for the considered scenario.}
    \label{fig:diff}
\end{figure}
\begin{figure}[t]
    \centering
    \includegraphics[width=0.6\columnwidth]{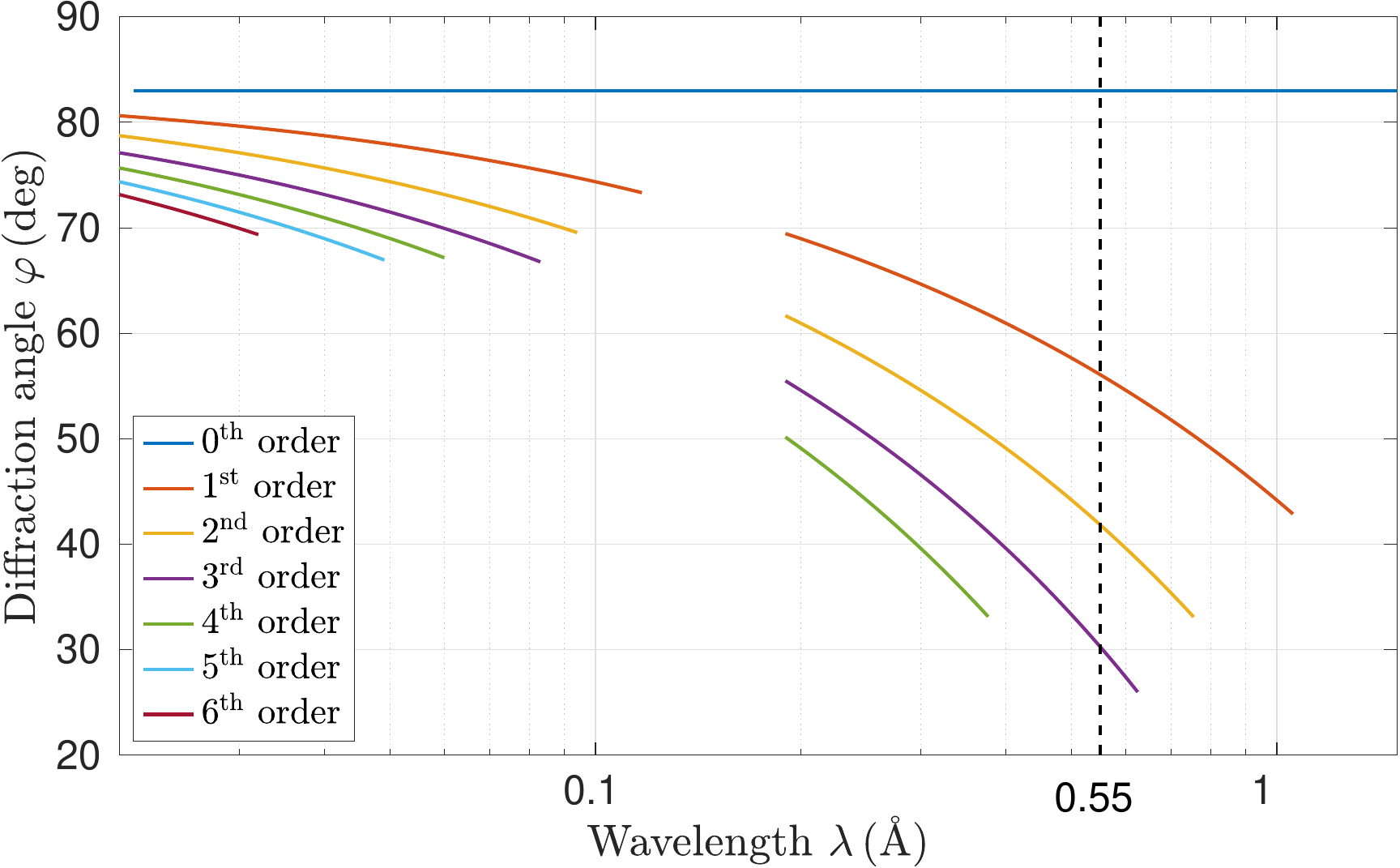}
    \caption{Positions of the diffraction orders [$0^{\rm th}$ (blue); $1^{\rm st}$ (orange); $2^{\rm nd}$ (yellow); $3^{\rm rd}$ (purple); $4^{\rm th}$ (green); $5^{\rm th}$ (light blue); $6^{\rm th}$ (dark red)] for an incidence angle of $83$ deg depending on the wavelength of the helium beam.}
    \label{fig:kdep50}
\end{figure}

Due to the separation of the length scales and the fact that the atoms interfere with themselves and not with each other, we describe each interference pattern via a phase-shifted Gaussian wave in analogy to the Michelson interferometer. Thus, the interference pattern is described by the superposition of phase-shifted Gaussian waves 
\begin{align}
I(\varphi) \propto \left|\sum_n a_n\me^{\mi \frac{k\sin\varphi}{2}b_n} \right|^2 \me^{-\frac{2 L_2^2\sin^2\varphi}{ w^2}}\,,\label{eq:intens}
\end{align}
with the amplitudes $a_n=\rho_{n_1}\rho_{n_2}\rho_{n_3}$ and the optical path lengths $b_n$, which are given in table~\ref{tab:diff}. The widths of the diffraction orders $\sigma$ are small compared to the width of the Gaussian envelope, $L_2\sin\sigma \ll w$, and, hence, can be neglected. It can be seen in Eq.~(\ref{eq:intens}) that the interference fringes are determined by the wave vector, $k=2\pi/\lambda_{\rm dB}$. Thus, increasing the wavelength, either by increasing the particle's mass or velocity, will reduce the spacing between the interference fringes. The resulting interference patterns are plotted in Fig.~\ref{fig:diff}. One can see that the diffraction at $30.32$ deg and $41.87$ deg will not show any interference features due to the equal optical path lengths of both optical paths. The remaining two spots will show interference effects with a contrast of 48.5\% for the spot at 56.10 deg and 84.1\% for the spot at 83.00 deg. The transmission rates of all channels can be found in table~\ref{tab:diff}: 0.027\% of the atoms will be diffracted under the angle of $30.32$ deg, 0.0675\% under $41.87$ deg, 0.1688\% under $56.10$ deg, 0.216\% under $83$ deg. The remaining particles will not leave the interferometer.  The intensity of a typical helium beam is so big~\cite{PhysRevA.98.063611} that a signal fraction of $10^{-4}$ can easily be detected. The velocity spread will be the limiting quantity to measure the interference patterns for the helium atom interferometry configuration depicted in Fig.~\ref{fig:diff}. The velocity spread of a supersonic helium beam depends on the beam temperature, the nozzle diameter and the reservoir pressure. This has been treated extensively in the literature; see, for example, Ref.~\cite{10.1063/1.5044203}. A velocity spread causes two different effects: (i) a broadening of the interference fringes and (ii) a spatial movement of the entire interference pattern, as illustrated in Fig.~\ref{fig:kdep50}. Finally, to observe interference, the velocity spread has to be sufficiently small to not cause a washing out of the interference fringes. Figure~\ref{fig:kdep50} illustrates the positions of the diffraction order for different wavelengths of the incoming beam with a fixed incidence angle of $83$ deg. It can be observed that the zeroth order will stay constant. The remaining orders strongly spread out with increasing wavelength. As in Fig.~\ref{fig:diffractionorders}, the lines are not continuous due to the finite size of the interferometer. It can be seen in table~\ref{tab:diff} that the interferometer splits the wave package at the first diffraction point over $\approx 0.71\,\rm{rad}$.

\subsection{Quantum reflection interferometer}
Quantum reflection occurs on the attractive (outer) part of the atom-surface interaction potential~\cite{D2CP03349F} in contrast to surface scattering, where the reflection occurs on the repulsive (inner) part of the interaction potential~\cite{PhysRevX.4.011029,Galiffi_2017}. It is called quantum reflection because, classically, reflection cannot occur with an attractive force interaction potential. Quantum reflection has the very big advantage that an extensive range of atoms and small, few-atomic molecules that would stick under surface diffraction conditions display quantum reflection.  The disadvantage is that quantum reflection requires small perpendicular wave vectors. This means that, for a given wavelength, the spatial extension of a  reflective interferometer  must be larger than in the surface scattering configuration for the separated beams to recombine.

Quantum reflection is less sensitive than surface scattering to defects and surface contamination because it occurs at larger distances from the surface~\cite{Stickler2014QuantumRA}.  Very large specular reflection coefficients of the order of 50\%~\cite{PhysRevA.78.010902} up to 90\%~\cite{Judd_2011} have been measured. Diffraction via quantum reflection was recently demonstrated experimentally~\cite{D2CP02641D} using helium dimers and trimers with periodically striped surfaces with micron-sized structures. The paper includes a comparison of the experimental result with scattering theory based on the diffraction angle distribution~(\ref{eq:diff}), reported in Ref.~\cite{EBogomolny2003}. There is reasonable agreement between theory and experiment. 

\section{Conclusions and Future Work}

This paper presents the first proposal for a reflective interferometer for atoms and molecules. We present calculations for a monolithic configuration based on experimental scattering results for a room-temperature helium beam from Si(111)-H(1$\times 1$), showing that a beam splitting of more than 0.5 radians is achievable. Furthermore, we argue that quantum reflection diffraction is a viable option for extending the beams and surfaces that can be used and potentially increase the signal intensity. The interference of larger and complex molecules can be achieved by using different interaction potentials, such as evanescent fields~\cite{PhysRevX.4.011029}. A reflective atom or molecule interferometer, particularly in a monolithic configuration, opens several possibilities for applications, for instance, as an accelerometer, in investigating the coherence of matters near dielectric surfaces,  as a continuous velocity selector etc. 
The next obvious first step is to do a demonstration experiment of the new interferometer with a helium beam and to do detailed designs of quantum reflection setups. The latter will require the calculation of quantum (diffraction) reflection coefficients for a range of realistic system configurations.

\section*{Acknowledgments}
J.F. gratefully acknowledges support from the European Union (H2020-MSCA-IF-2020, grant number: 101031712).

%Bibliography
\bibliographystyle{unsrt}  
\bibliography{apssamp}

\providecommand{\noopsort}[1]{}\providecommand{\singleletter}[1]{#1}%\providecommand{\noopsort}[1]{}\providecommand{\singleletter}[1]{#1}%
\begin{thebibliography}{10}

\bibitem{stray2022quantum}
Ben Stray, Andrew Lamb, Aisha Kaushik, Jamie Vovrosh, Anthony Rodgers, Jonathan
  Winch, Farzad Hayati, Daniel Boddice, Artur Stabrawa, Alexander Niggebaum,
  Mehdi Langlois, Yu-Hung Lien, Samuel Lellouch, Sanaz Roshanmanesh, Kevin
  Ridley, Geoffrey de~Villiers, Gareth Brown, Trevor Cross, George Tuckwell,
  Asaad Faramarzi, Nicole Metje, Kai Bongs, and Michael Holynski.
\newblock Quantum sensing for gravity cartography.
\newblock {\em Nature}, 602(7898):590--594, 2022.

\bibitem{hardman2016simultaneous}
K.~S. Hardman, P.~J. Everitt, G.~D. McDonald, P.~Manju, P.~B. Wigley, M.~A.
  Sooriyabandara, C.~C.~N. Kuhn, J.~E. Debs, J.~D. Close, and N.~P. Robins.
\newblock Simultaneous precision gravimetry and magnetic gradiometry with a
  bose-einstein condensate: A high precision, quantum sensor.
\newblock {\em Physical Review Letters}, 117(13):138501, Sep 2016.

\bibitem{riedel2010atom-chip-based}
Max~F. Riedel, Pascal B{\"o}hi, Yun Li, Theodor~W. H{\"a}nsch, Alice Sinatra,
  and Philipp Treutlein.
\newblock Atom-chip-based generation of entanglement for quantum metrology.
\newblock {\em Nature}, 464(7292):1170--1173, 2010.

\bibitem{ludlow2015optical}
Andrew~D. Ludlow, Martin~M. Boyd, Jun Ye, E.~Peik, and P.O. Schmidt.
\newblock Optical atomic clocks.
\newblock {\em Reviews of Modern Physics}, 87(2):637--701, jun 2015.

\bibitem{canuel2020elgar}
Benjamin Canuel, S~Abend, P~Amaro-Seoane, F~Badaracco, Q~Beaufils, Andrea
  Bertoldi, Kai Bongs, Philippe Bouyer, Claus Braxmaier, W~Chaibi,
  N~Christensen, F~Fitzek, G~Flouris, Naceur Gaaloul, S~Gaffet, C~L~Garrido
  Alzar, R~Geiger, S~Guellati-Khelifa, Klemens Hammerer, J~Harms, J~Hinderer,
  M~Holynski, J~Junca, S~Katsanevas, C~Klempt, C~Kozanitis, Markus Krutzik,
  Arnaud Landragin, I~L{\`{a}}zaro Roche, B~Leykauf, Y-H Lien, S~Loriani,
  S~Merlet, M~Merzougui, M~Nofrarias, P~Papadakos, F~Pereira dos Santos, Achim
  Peters, D~Plexousakis, M~Prevedelli, E~M Rasel, Y~Rogister, S~Rosat, Albert
  Roura, D~O Sabulsky, V~Schkolnik, D~Schlippert, Christian Schubert,
  L~Sidorenkov, J-N Siem{\ss}, C~F Sopuerta, Fiodor Sorrentino, C~Struckmann,
  Guglielmo~Maria Tino, G~Tsagkatakis, A~Vicer{\'{e}}, Wolf von Klitzing,
  L~Woerner, and X~Zou.
\newblock Elgar -- a european laboratory for gravitation and
  atom-interferometric research.
\newblock {\em Classical and Quantum Gravity}, 37(22):225017, oct 2020.

\bibitem{el-neaj2020aedge}
Yousef~Abou El-Neaj, Cristiano Alpigiani, Sana Amairi-Pyka, Henrique
  Ara{\'u}jo, Antun Bala{\v z}, Angelo Bassi, Lars Bathe-Peters, Baptiste
  Battelier, Aleksandar Beli{\'c}, Elliot Bentine, Jos{\'e} Bernabeu, Andrea
  Bertoldi, Robert Bingham, Diego Blas, Vasiliki Bolpasi, Kai Bongs, Sougato
  Bose, Philippe Bouyer, Themis Bowcock, William Bowden, Oliver Buchmueller,
  Clare Burrage, Xavier Calmet, Benjamin Canuel, Laurentiu-Ioan Caramete,
  Andrew Carroll, Giancarlo Cella, Vassilis Charmandaris, Swapan Chattopadhyay,
  Xuzong Chen, Maria~Luisa Chiofalo, Jonathon Coleman, Joseph Cotter, Yanou
  Cui, Andrei Derevianko, Albert De~Roeck, Goran~S. Djordjevic, Peter Dornan,
  Michael Doser, Ioannis Drougkakis, Jacob Dunningham, Ioana Dutan, Sajan Easo,
  Gedminas Elertas, John Ellis, Mai El~Sawy, Farida Fassi, Daniel Felea,
  Chen-Hao Feng, Robert Flack, Chris Foot, Ivette Fuentes, Naceur Gaaloul,
  Alexandre Gauguet, Remi Geiger, Valerie Gibson, Gian Giudice, Jon Goldwin,
  Oleg Grachov, Peter~W. Graham, Dario Grasso, Maurits van~der Grinten, Mustafa
  G{\"u}ndogan, Martin~G. Haehnelt, Tiffany Harte, Aur{\'e}lien Hees, Richard
  Hobson, Jason Hogan, Bodil Holst, Michael Holynski, Mark Kasevich, Bradley~J.
  Kavanagh, Wolf von Klitzing, Tim Kovachy, Benjamin Krikler, Markus Krutzik,
  Marek Lewicki, Yu-Hung Lien, Miaoyuan Liu, Giuseppe~Gaetano Luciano, Alain
  Magnon, Mohammed~Attia Mahmoud, Sarah Malik, Christopher McCabe, Jeremiah
  Mitchell, Julia Pahl, Debapriya Pal, Saurabh Pandey, D.~G. Papazoglou, Mauro
  Paternostro, Bjoern Penning, Achim Peters, Marco Prevedelli, Vishnupriya
  Puthiya-Veettil, John Quenby, Ernst Rasel, Sean Ravenhall, Jack Ringwood,
  Albert Roura, D~O Sabulsky, Muhammed Sameed, Ben Sauer, Stefan~Alaric
  Sch{\"a}ffer, Stephan Schiller, Vladimir Schkolnik, Dennis Schlippert,
  Christian Schubert, Haifa~Rejeb Sfar, Armin Shayeghi, Ian Shipsey, Carla
  Signorini, Yeshpal Singh, Marcelle Soares-Santos, Fiodor Sorrentino, Timothy
  Sumner, Konstantinos Tassis, Silvia Tentindo, Guglielmo~Maria Tino,
  Jonathan~N. Tinsley, James Unwin, Tristan Valenzuela, Georgios Vasilakis,
  Ville Vaskonen, Christian Vogt, Alex Webber-Date, Andr{\'e} Wenzlawski,
  Patrick Windpassinger, Marian Woltmann, Efe Yazgan, Ming-Sheng Zhan, Xinhao
  Zou, and Jure Zupan.
\newblock Aedge - atomic experiment for dark matter and gravity exploration in
  space.
\newblock {\em EPJ Quantum Technology}, 7(1):6, 2020.

\bibitem{Tino2013}
G.~M. Tino and et~al.
\newblock Precision gravity tests with atom interferometry in space.
\newblock {\em Nuc. Phys. B}, 243-244:203, 2013.

\bibitem{desruelle2018}
V.~Menoret, P.~Vermeulen, N.~Le Mogine, S.~Bonvalot, P.~Bouyer, A.~Landragin,
  and B.~Desruelle.
\newblock Gravity measurements below $10^{-9}~g$ with a transportable absolute
  quantum gravimeter.
\newblock {\em Scientific Reports}, 8:12300, 2018.

\bibitem{app10041256}
Josué González-García, Alfonso Gómez-Espinosa, Enrique Cuan-Urquizo,
  Luis~Govinda García-Valdovinos, Tomás Salgado-Jiménez, and Jesús
  Arturo~Escobedo Cabello.
\newblock Autonomous underwater vehicles: Localization, navigation, and
  communication for collaborative missions.
\newblock {\em Applied Sciences}, 10(4), 2020.

\bibitem{hall1998measurements}
D.~S. Hall, M.~R. Matthews, C.~E. Wieman, and E.~A. Cornell.
\newblock Measurements of relative phase in two-component bose-einstein
  condensates.
\newblock {\em Physical Review Letters}, 81(8):1543--1546, 3 1998.

\bibitem{Pritchard1991}
D.~W. Keith, C.~R. Ekstrom, Q.~A. Turchette, and D.~E. Pritchard.
\newblock An interferometer for atoms.
\newblock {\em Physical Review Letters}, 66(21):2693--2696, 3 1991.

\bibitem{biedermann2017atom}
G.~W. Biedermann, H.J. McGuinness, A.{\hspace{0.167em}}V. Rakholia, Y.-Y. Jau,
  D.R. Wheeler, J.D. Sterk, and G.R. Burns.
\newblock Atom interferometry in a warm vapor.
\newblock {\em Physical Review Letters}, 118(16), apr 2017.

\bibitem{Carnal1991}
O~Carnal, A~Faulstich, and J~Mlynek.
\newblock {Diffraction of metastable helium atoms by a transmission grating}.
\newblock {\em Applied Physics B}, 53(2):88--91, 1991.

\bibitem{Lee1995}
D.~M. Giltner, R.~W. McGowan, and S.~A. Lee.
\newblock Atom interferometer based on bragg scattering from standing light
  waves.
\newblock {\em Phys. Rev. Lett.}, 75:2638, 1995.

\bibitem{Zeilinger1995}
E.~M. Rasel, M.~K. Oberthaler, H.~Batelaan, J.~Schmiedmayer, and A.~Zeilinger.
\newblock Atom wave interferometry with diffraction gratings of light.
\newblock {\em Phys. Rev. Lett.}, 75:2633, 1995.

\bibitem{Vigue2002}
R.~Delhuille, C.~Champenois, M.~Buchner, L.~Jozefowski, C.~Rizzo, G.~Trénec,
  and J.~Vigué.
\newblock High-contrast mach–zehnder lithium-atom interferometer in the bragg
  regime.
\newblock {\em Appl. Phys. B}, 74:489, 2002.

\bibitem{Pritchard2009}
A.~D. Cronin.
\newblock Optics and interferometry with atoms and molecules.
\newblock {\em Rev. Mod. Phys}, 81:1051, 2009.

\bibitem{Li94}
J.~F. Clauser and S:~Li.
\newblock Talbov-vonlau atom interferometry with cold slow potassium.
\newblock {\em Phys. Rev. A}, 49:R2213, 1994.

\bibitem{PhysRev.78.695}
Norman~F. Ramsey.
\newblock A molecular beam resonance method with separated oscillating fields.
\newblock {\em Phys. Rev.}, 78:695--699, Jun 1950.

\bibitem{PhysRevA.30.1836}
Ch.~J. Bord\'e, Ch. Salomon, S.~Avrillier, A.~van Lerberghe, Ch. Br\'eant,
  D.~Bassi, and G.~Scoles.
\newblock Optical ramsey fringes with traveling waves.
\newblock {\em Phys. Rev. A}, 30:1836--1848, Oct 1984.

\bibitem{Riehle91}
F.~Riehle, Th. Kisters, A.~Witte, J.~Helmcke, and Ch.~J. Bord\'e.
\newblock Optical ramsey spectroscopy in a rotating frame: Sagnac effect in a
  matter-wave interferometer.
\newblock {\em Phys. Rev. Lett.}, 67:177--180, Jul 1991.

\bibitem{doi:10.1063/1.1754212}
U.~Bonse and M.~Hart.
\newblock An x‐ray interferometer.
\newblock {\em Applied Physics Letters}, 6(8):155--156, 1965.

\bibitem{Chetwynd_1991}
D~G Chetwynd, S~C Cockerton, S~T Smith, and W~W Fung.
\newblock The design and operation of monolithic x-ray interferometers for
  super-precision metrology.
\newblock {\em Nanotechnology}, 2(1):1, jan 1991.

\bibitem{Rauch2015}
H.~Rauch and S.~A. Werner.
\newblock {\em Neutron Interferometry}.
\newblock Oxford University Press, 2015.

\bibitem{berggren2017}
Akshay Agarwal, Chung-Soo Kim, Richard Hobbs, Dirk van Dyck, and Karl~K.
  Berggren.
\newblock A nanofabricated, monolithic, path-separated electron interferometer.
\newblock {\em Scientific Reports}, 7(1):1677, may 2017.

\bibitem{PhysRevLett.34.1472}
R.~Colella, A.~W. Overhauser, and S.~A. Werner.
\newblock Observation of gravitationally induced quantum interference.
\newblock {\em Phys. Rev. Lett.}, 34:1472--1474, Jun 1975.

\bibitem{Nesvizhevsky2002}
Valery~V Nesvizhevsky, Hans~G B{\"{o}}rner, Alexander~K Petukhov, Hartmut
  Abele, Stefan Bae{\ss}ler, Frank~J Rue{\ss}, Thilo St{\"{o}}ferle, Alexander
  Westphal, Alexei~M Gagarski, Guennady~A Petrov, and Alexander~V Strelkov.
\newblock {Quantum states of neutrons in the Earth's gravitational field}.
\newblock {\em Nature}, 415(6869):297--299, 2002.

\bibitem{PhysRevLett.115.013004}
Lin Zhou, Shitong Long, Biao Tang, Xi~Chen, Fen Gao, Wencui Peng, Weitao Duan,
  Jiaqi Zhong, Zongyuan Xiong, Jin Wang, Yuanzhong Zhang, and Mingsheng Zhan.
\newblock Test of equivalence principle at $1{0}^{\ensuremath{-}8}$ level by a
  dual-species double-diffraction raman atom interferometer.
\newblock {\em Phys. Rev. Lett.}, 115:013004, Jul 2015.

\bibitem{PhysRevLett.78.2046}
T.~L. Gustavson, P.~Bouyer, and M.~A. Kasevich.
\newblock Precision rotation measurements with an atom interferometer
  gyroscope.
\newblock {\em Phys. Rev. Lett.}, 78:2046--2049, Mar 1997.

\bibitem{doi:10.1080/09500349708231906}
Herman Batelaan, Ernst~M. Rasel, Markus~K. Oberthaler, Jörg Schmiedmayer, and
  Anton Zeilinger.
\newblock Anomalous transmission in atom optics.
\newblock {\em Journal of Modern Optics}, 44(11-12):2629--2641, 1997.

\bibitem{PhysRevLett.66.2693}
David~W. Keith, Christopher~R. Ekstrom, Quentin~A. Turchette, and David~E.
  Pritchard.
\newblock An interferometer for atoms.
\newblock {\em Phys. Rev. Lett.}, 66:2693--2696, May 1991.

\bibitem{PhysRevA.49.R2213}
John~F. Clauser and Shifang Li.
\newblock Talbot-vonlau atom interferometry with cold slow potassium.
\newblock {\em Phys. Rev. A}, 49:R2213--R2216, Apr 1994.

\bibitem{PhysRevLett.127.170402}
C.~Garcion, N.~Fabre, H.~Bricha, F.~Perales, S.~Scheel, M.~Ducloy, and
  G.~Dutier.
\newblock Intermediate-range casimir-polder interaction probed by high-order
  slow atom diffraction.
\newblock {\em Phys. Rev. Lett.}, 127:170402, Oct 2021.

\bibitem{doi:10.1080/01442350500037521}
Alec~M. Wodtke, John~C. Tully, and Daniel~J. Auerbach.
\newblock Electronically non-adiabatic interactions of molecules at metal
  surfaces: Can we trust the born–oppenheimer approximation for surface
  chemistry?
\newblock {\em International Reviews in Physical Chemistry}, 23(4):513--539,
  2004.

\bibitem{doi:10.1146/annurev-physchem-040214-121958}
Kai Golibrzuch, Nils Bartels, Daniel~J. Auerbach, and Alec~M. Wodtke.
\newblock The dynamics of molecular interactions and chemical reactions at
  metal surfaces: Testing the foundations of theory.
\newblock {\em Annual Review of Physical Chemistry}, 66(1):399--425, 2015.
\newblock PMID: 25580627.

\bibitem{saxena1989thermal}
S.C. Saxena and R.K. Joshi.
\newblock {\em Thermal Accommodation and Adsorption Coefficients of Gases}.
\newblock CINDAS data series on material properties. Hemisphere Publishing
  Corporation, 1989.

\bibitem{D2CP03369K}
Arved~C. Dorst, Friedrich Güthoff, Daniel Schauermann, Alec~M. Wodtke,
  Daniel~R. Killelea, and Tim Schäfer.
\newblock Velocity map images of desorbing oxygen from sub-surface states of
  rh(111).
\newblock {\em Phys. Chem. Chem. Phys.}, pages~--, 2022.

\bibitem{doi:10.1063/5.0058789}
Loïc Lecroart, Nils Hertl, Yvonne Dorenkamp, Hongyan Jiang, Theofanis~N.
  Kitsopoulos, Alexander Kandratsenka, Oliver Bünermann, and Alec~M. Wodtke.
\newblock Adsorbate modification of electronic nonadiabaticity: H atom
  scattering from p(2 × 2) o on pt(111).
\newblock {\em The Journal of Chemical Physics}, 155(3):034702, 2021.

\bibitem{doi:10.1063/5.0026228}
Jan Geweke and Alec~M. Wodtke.
\newblock Vibrationally inelastic scattering of hcl from ag(111).
\newblock {\em The Journal of Chemical Physics}, 153(16):164703, 2020.

\bibitem{palau2021}
A.~S. Palau, S.~D. Eder, G.~Bracco, and B.~Holst.
\newblock Neutral helium atom microscopy.
\newblock {\em https://arxiv.org/pdf/2111.12582.pdf}, 2021.

\bibitem{holst1997}
B.~Holst and W.~Allison.
\newblock An atom-focusing mirror.
\newblock {\em Nature}, 8:244, 1997.

\bibitem{BARREDO200724}
D.~Barredo, F.~Calleja, A.E. Weeks, P.~Nieto, J.J. Hinarejos, G.~Laurent, A.L.
  {Vazquez de Parga}, D.A. MacLaren, D.~Farías, W.~Allison, and R.~Miranda.
\newblock {S}i(111)–{H}(1×1): A mirror for atoms characterized by {AFM},
  {STM}, {H}e and {H}$_2$ diffraction.
\newblock {\em Surface Science}, 601(1):24--29, 2007.

\bibitem{doi:10.1063/1.480723}
J.~R. Buckland and W.~Allison.
\newblock Determination of the helium/{S}i(111)–(1×1){H} potential.
\newblock {\em The Journal of Chemical Physics}, 112(2):970--978, 2000.

\bibitem{Buckland1999}
J.~R. Buckland, B.~Holst, and W.~Allison.
\newblock Helium reflectivity of the {S}i(111)-{H}(1$\times$1) surface for use
  in atom optical elements.
\newblock {\em Chem. Phys. Lett.}, 303:107--110, 199.

\bibitem{MACLAREN2001285}
D.A. MacLaren, N.J. Curson, P.~Atkinson, and W.~Allison.
\newblock An {AFM} study of the processing of hydrogen passivated silicon(111)
  of a low miscut angle.
\newblock {\em Surface Science}, 490(3):285--295, 2001.

\bibitem{Brand2015}
Christian Brand, Michele Sclafani, Christian Knobloch, Yigal Lilach, Thomas
  Juffmann, Jani Kotakoski, Clemens Mangler, Andreas Winter, Andrey Turchanin,
  Jannik Meyer, Ori Cheshnovsky, and Markus Arndt.
\newblock {An atomically thin matter-wave beamsplitter}.
\newblock {\em Nature Nanotechnology}, 10(10):845--848, 2015.

\bibitem{PhysRevLett.125.033604}
Christian Brand, Filip Kia\l{}ka, Stephan Troyer, Christian Knobloch, Ksenija
  Simonovi\ifmmode~\acute{c}\else \'{c}\fi{}, Benjamin~A. Stickler, Klaus
  Hornberger, and Markus Arndt.
\newblock Bragg diffraction of large organic molecules.
\newblock {\em Phys. Rev. Lett.}, 125:033604, Jul 2020.

\bibitem{Bruhl_2002}
R.~Brühl, P.~Fouquet, R.~E. Grisenti, J.~P. Toennies, G.~C. Hegerfeldt,
  T.~Köhler, M.~Stoll, and C.~Walter.
\newblock The van der waals potential between metastable atoms and solid
  surfaces: Novel diffraction experiments vs. theory.
\newblock {\em Europhysics Letters}, 59(3):357, aug 2002.

\bibitem{Kittel2004}
Charles Kittel.
\newblock {\em Introduction to Solid State Physics}.
\newblock Wiley, 8 edition, 2004.

\bibitem{PhysRevA.98.063611}
Adri\`a~Salvador Palau, Sabrina~D. Eder, Truls Andersen, Anders~Kom\'ar Ravn,
  Gianangelo Bracco, and Bodil Holst.
\newblock Center-line intensity of a supersonic helium beam.
\newblock {\em Phys. Rev. A}, 98:063611, Dec 2018.

\bibitem{10.1063/1.5044203}
S.~D. Eder, A.~Salvador~Palau, T.~Kaltenbacher, G.~Bracco, and B.~Holst.
\newblock {Velocity distributions in microskimmer supersonic expansion helium
  beams: High precision measurements and modeling}.
\newblock {\em Review of Scientific Instruments}, 89(11), 11 2018.
\newblock 113301.

\bibitem{D2CP03349F}
J.~Fiedler, K.~Berland, J.~W. Borchert, R.~W. Corkery, A.~Eisfeld,
  D.~Gelbwaser-Klimovsky, M.~M. Greve, B.~Holst, K.~Jacobs, M.~Krüger, D.~F.
  Parsons, C.~Persson, M.~Presselt, T.~Reisinger, S.~Scheel, F.~Stienkemeier,
  M.~Tømterud, M.~Walter, R.~T. Weitz, and J.~Zalieckas.
\newblock Perspectives on weak interactions in complex materials at different
  length scales.
\newblock {\em Phys. Chem. Chem. Phys.}, 25:2671--2705, 2023.

\bibitem{PhysRevX.4.011029}
Helmar Bender, Christian Stehle, Claus Zimmermann, Sebastian Slama, Johannes
  Fiedler, Stefan Scheel, Stefan~Yoshi Buhmann, and Valery~N. Marachevsky.
\newblock Probing atom-surface interactions by diffraction of bose-einstein
  condensates.
\newblock {\em Phys. Rev. X}, 4:011029, Feb 2014.

\bibitem{Galiffi_2017}
Emanuele Galiffi, Christoph Sünderhauf, Maarten DeKieviet, and Sandro
  Wimberger.
\newblock Two-dimensional simulation of quantum reflection.
\newblock {\em Journal of Physics B: Atomic, Molecular and Optical Physics},
  50(9):095001, apr 2017.

\bibitem{Stickler2014QuantumRA}
Benjamin~A. Stickler, Uzi Even, and Klaus Hornberger.
\newblock Quantum reflection and interference of matter waves from periodically
  doped surfaces.
\newblock {\em Physical Review A}, 91:013614, 2014.

\bibitem{PhysRevA.78.010902}
Bum~Suk Zhao, Stephan~A. Schulz, Samuel~A. Meek, Gerard Meijer, and Wieland
  Sch\"ollkopf.
\newblock Quantum reflection of helium atom beams from a microstructured
  grating.
\newblock {\em Phys. Rev. A}, 78:010902, Jul 2008.

\bibitem{Judd_2011}
T~E Judd, R~G Scott, A~M Martin, B~Kaczmarek, and T~M Fromhold.
\newblock Quantum reflection of ultracold atoms from thin films, graphene and
  semiconductor heterostructures.
\newblock {\em New Journal of Physics}, 13(8):083020, aug 2011.

\bibitem{D2CP02641D}
Lee~Yeong Kim, Sanghwan Park, Chang~Young Lee, Wieland Schöllkopf, and Bum~Suk
  Zhao.
\newblock Enhanced elastic scattering of he2 and he3 from solids by
  multiple-edge diffraction.
\newblock {\em Phys. Chem. Chem. Phys.}, 24:21593--21600, 2022.

\bibitem{EBogomolny2003}
E~Bogomolny and C~Schmit.
\newblock Asymptotic behaviour of multiple scattering on an infinite number of
  parallel half-planes.
\newblock {\em Nonlinearity}, 16(6):2035, sep 2003.

\end{thebibliography}

\end{document}